# Elimination of degenerate trajectory of single atom strongly coupled to the tilted cavity $TEM_{10}$ mode

Pengfei Zhang, Yanqiang Guo, Zhuoheng Li, Yanfeng Zhang, Jinjin Du, Gang Li, Junmin Wang and Tiancai Zhang*

*State Key Laboratory of Quantum Optics and Quantum Optics Devices, Institute of Opto-Electronics, Shanxi University, Taiyuan 030006, China*

**Abstract:** We demonstrate the trajectory measurement of the single neutral atoms deterministically using a high-finesse optical micro-cavity. Single atom strongly couples to the high-order transverse vacuum $TEM_{10}$ mode, instead of the usual $TEM_{00}$ mode, and the parameter of the system is $(g_{10}, \kappa, \gamma)/2\pi = (20.5, 2.6, 2.6) MHz$. The atoms simply fall down freely from the magneto-optic trap into the cavity modes and the trajectories of the single atoms are linear. The transmission spectrums of atoms passing through the $TEM_{10}$ mode are detected by a single photon counting modules and well fitted. Thanks to the tilted cavity transverse $TEM_{10}$ mode, which is inclined to the vertical direction about 45 degrees and it helps us, for the first time, to eliminate the degenerate trajectory of the single atom falling through the cavity and get the unique atom trajectory. Atom position with high precision of 0.1μm in the off-axis direction (axis *y*) is obtained, and the spatial resolution of 5.6μm is achieved in time of 10μs along the vertical direction (axis *x*). The average velocity of the atoms is also measured from the atom transits, which determines the temperature of the atoms in magneto-optic trap, 186μK±19μK.

PACS number(s): 42.50.Pq, 37.10.Jk

Manipulation of neutral single atoms, known as the basic system of quantum optics and atomic physics, has been extensively studied since the last two decades, either in free space [1-6] or inside a cavity [7-10]. From the early hot and stochastic atom beam to the cold and deterministic control of individual atoms, single atom is now a good system to demonstrate quantum logic gate [11] and quantum register [4,

12]. In order to get information from a single atom two predominant methods are used in most experiments. One is to detect directly the fluorescence of the atoms using a special designed light collection system and high-efficiency optical detector; another is to detect the transmitted light from a high-finesse optical cavity, which is usually strongly coupled to the atoms [13]. The cavity quantum electrodynamics (CQED) system has been used to detect single atoms as well as the atom trajectory [14]. Large coupling between single atom and cavity provides the capability of measuring the atomic trajectory through the transmission of the cavity [15, 16].

In the earlier experiments, atom beam has been used in CQED experiments [17-19] and the duration of the atom transits was so short that the detection of individual atom positions and its trajectory could not be accomplished in real time. The development of the cold atoms technology and the manipulation of single atoms [20, 21] provide the effective tools for the CQED experiments. Either through the atom free falling down or launching up to the cavity, the transit time of atoms in the cavity mode lasted more than 100μs [15, 16] and the trajectories of single atom can be measured. In 2000, *Hood et al* trapped single atom inside a micro-cavity for milliseconds and the 2D atom trajectories in the plane perpendicular to the cavity axis were reconstructed from the cavity transmissions and they obtained 2μm of the spatial resolution in a 10μs of time interval [14]. But for all these experiments mentioned above, the atom was coupled to the fundamental Hermite-Gaussian $TEM_{00}$ mode. Although the coupling between atom and cavity $TEM_{00}$ mode is stronger than all the other modes when the atom passes in the centre of cavity mode, the displacements of the atom along the cavity axis, or between a node and antinode, can not be determined since the spatial symmetry of the $TEM_{00}$ mode causes the quadruple degeneracy of the atom trajectories in principle. Higher order transverse modes may break the spatial symmetry and reduce the degeneracy of the atom trajectories. In 2003, *Puppe et al.* demonstrated the single-atom trajectories in high-order transverse modes of a high-finesse optical cavity [22]. The atom trajectories were obtained according to the transmission spectrum of the cavity. However, the spatial patterns of $TEM_{01}$ and $TEM_{10}$ modes were oriented nearly horizontally and the atom trajectories were still degenerate, but from quadruple degeneracy to duplicate degeneracy. In this letter, a tilted spatial transverse TEM$_{10}$ mode is used, which breaks the symmetry and allows eliminating the degeneracy of the atom trajectories completely. We use the strong coupled atom-cavity system to track the atomic path and determine the ballistic trajectory of a single atom uniquely. The spatial resolution of 5.6μm is achieved in the time of 10μs along the vertical direction (axis x), while the atom position along the horizontal direction (axis y) can be obtained with precision of 0.1μm. With the help of even higher order modes and smaller mode waist of the high-finesse cavity, it is capable to obtain the trajectory of the single atom with high spatial resolution by the so-called atomic kaleidoscope [23, 24] based on the atom-cavity microscope (ACM) system [14].

The CQED system contains a cavity and an atom couples to a single mode of the electromagnetic field as shown in Fig. 1. The interaction between the atom and the single-mode field is described by the oscillatory exchange of energy (Rabi Oscillation), characterized by *g*. Real experimental system is an open system and both of cavity decay rate *κ* and atom decay rate *γ* must be taken in account. In the strong coupling regimes the optimum coupling constant g$_0$ in the TEM$_{00}$ mode is much lager than *κ* and *γ*. The cavity transmission in the weak-field limit of small excited state population is [25]

$$T(x,y,z) = \frac{\kappa^2(\gamma^2 + \Delta_{pa}^2)}{\left[g_{eff}(x,y,z)^2 - \Delta_{pa}^2 - \Delta_{ca}\Delta_{pa} + \gamma\kappa\right]^2 + (\kappa\Delta_{pa} + \gamma\Delta_{pa} - \gamma\Delta_{ca})^2}. \quad (1)$$

Here $\Delta_{pa} = \omega_{probe} - \omega_{atom}$ is the detuning between probe light and atomic transition, and $\Delta_{ca} = \omega_{cavity} - \omega_{atom}$ is the detuning between the cavity and the atom. The effective coupling *g$_{eff}$* for arbitrary Hermite-Gaussian transverse modes is dependent on the position of the atom in the cavity, $g_{eff}(x,y,z) = g_0 \Psi_{m,n}(x,y,z)/\Psi_{0,0}(0,0,0)$. Here we have assumed the cavity axis is z and $\Psi_{m,n}(x,y,z)$ are the mode functions of the cavity, expressed by

$$\Psi_{m,n}(x,y,z) = C_{m,n} \exp(-\frac{x^2+y^2}{w_0^2}) H_m(\frac{\sqrt{2}x}{w_0}) H_n(\frac{\sqrt{2}y}{w_0}) \cos(\frac{2\pi z}{\lambda}), \qquad (2)$$

Here $C_{m,n} = (2^m 2^n m! n!)^{-1/2} (w_0^2 \pi / 2)^{-1/2}$, and $H_{m,n}$ are the corresponding Hermite polynomials of order $m$ and $n$. $w_0$ is the waist of the mode which is determined by the radius of curvature of mirrors and the cavity length. $\lambda$ is the wavelength. The transmission given by (1) is position-dependent, thus from the spectrum of the cavity transmission, the atomic trajectory may be determined.

A conventional magneto-optic trap is positioned right above the cavity. The atoms fall down and pass through the cavity in gravitation when the MOT is shut off. Without trap inside the cavity, the single-atom trajectories of free falling atoms into the cavity from above are straight lines, as is physically appropriate. The position of atom along the cavity axis $z$ plays an important role for the transmission spectrum. However in the real experiments, since the cavity center is far from the MOT, the intra-cavity atom velocity in the horizontal direction is very small compare to that of the vertical direction, and the motion along the cavity axis $z$, which is most likely caused by the force due to the weak probe beam, is small [14]. Those transits corresponding to the transmissions dropping to nearly zero mean that the atoms pass through the antinode in a very narrow range in the cavity axis direction. The measured transmission spectrums will be simplified to the two-dimensional expression near the original of the antinode ($z \approx 0$):

$$T(x,y) = \frac{\kappa^2(\gamma^2 + \Delta_{pa}^2)}{\left[g_{eff}(x,y)^2 - \Delta_{pa}^2 - \Delta_{ca}\Delta_{pa} + \gamma\kappa\right]^2 + (\kappa\Delta_{pa} + \gamma\Delta_{pa} - \gamma\Delta_{ca})^2}, \qquad (3)$$

and the mode functions are

$$\Psi_{m,n}(x,y) = C_{m,n} \exp(-\frac{x^2+y^2}{w_0^2}) H_m(\frac{\sqrt{2}x}{w_0}) H_n(\frac{\sqrt{2}y}{w_0}). \qquad (4)$$

Here we have assumed that the trajectories of single-atom are linear. According to the expressions (3) and (4) we can simulate the trajectories of single-atom. Figure 2 shows a typical transmission spectrum for TEM$_{10}$ mode. Here we have used the parameters according to our experimental system: $g_0$=2π×23.9MHz, $w_0$=23.8μm,

$\gamma=2\pi\times2.6$MHz and $\kappa=2\pi\times2.6$MHz. The nodal lines of TEM$_{10}$ mode are assumed horizontally along the y axis. From Fig. 2 (a) we can see the double-dip structure appears when atom passed through the TEM$_{10}$ mode due to the intensity distribution of two pieces separated by a nodal line. The double-dip becomes smaller as the atom trajectory is getting far away from the center of mode (y is increasing from y=±20μm to y=±45μm). The cavity transmission spectrums are shown in Fig. 2(b) with different detunings between the probe and the atom transition for y=0. Double peaks appear with detuning $\Delta_{pa}$=-2π×23.9MHz (dashed-dotted green curve). A more complex spectrum will be seen for detuning $\Delta_{pa}$=-2π×10MHz (dotted blue curve). The dashed black curve is the effective coupling $g_{eff}$ in Fig. 2.

The experimental setup contains two important parts including the cold atom cloud trapped by MOT [26] and the high-finesse Fabry-Perot cavity composed by two spherical mirrors [27, 28]. Both of them are located in an ultrahigh vacuum chamber. The MOT is located 5mm above the cavity with the atom number of about $10^5$. The length of cavity is about 86μm with the waist of TEM$_{00}$ mode of $w_0$=23.8μm. The finesse of cavity is F=330000 and the parameters of the system are ($g_{10}$, $\kappa$, $\gamma$) = 2π×(20.5, 2.6, 2.6)MHz. The intra-cavity mean photon number is m≈1. The cavity is locked to the cesium D2 transition $6^2S_{1/2}, F=4 \to 6^2P_{3/2}, F'=5$. A single photon counting modules (SPCMs) [29] is used to detect the cavity transmission. Fig. 3 shows the cavity transmission versus the scanning of the frequency of probe light. The inset in Fig. 3 shows the TEM$_{00}$ and TEM$_{10/01}$ modes, respectively. In ideal case the transverse modes with the same *m+n* should be degenerate, but practical cavity gives a splitting about 82.5MHz between TEM$_{10}$ and TEM$_{01}$ modes, which is due to the geometric imperfection of the mirrors [27, 28]. The corresponding two dimensional distributions of the intensities of the modes are also obtained by a CCD camera and the patterns of the modes help us to determine directly the angle between the nodal lines of TEM$_{10/01}$ modes and *x* axis, which is about 45°.

The cavity transmission spectrums versus time with single atom passing through the cavity TEM$_{10}$ mode are shown in figure 4. Here the probe is resonant to the cavity.

The red dots are the experimental data and the green solid curves are the theoretical fittings according to the expression (3). There is an angle of $\theta=45°$ between the nodal lines of TEM$_{10}$ modes and x axis, and a rotating coordinate transformation is thus used: $x = x'\cos\theta + y'\sin\theta, y = -x'\sin\theta + y'\cos\theta$. It is clear to see from Fig. 4 that different double peaks of the transmission spectrums are observed, depending on the position of the atoms. The two dips are symmetrical when the atom trajectory is located at y=0±0.2μm (see Fig. 4 (b) and Fig. 4 (e)). The fitted atom velocity is 0.39±0.01m/s. When the atom trajectory is not at y=0, whatever for y<0 or y>0, the two dips are asymmetrical. For y<0, the former dip is deeper and wider and it is just the reverse for y>0. Thus the atom trajectory degeneracy is eliminated. Fig.4 (a) and (c) show the transmission spectrums corresponding to y=-16.3μm±0.1μm and y=+18.0μm±0.1μm, a left side transit and a right side transit, respectively. The spatial resolution in vertical direction (axis *x*) is about 5.6μm in 10μs time interval while 0.1μm for off-axis (*y* direction). The resolution for off-axis is much better than vertical direction, also much better than the reported result in Hood's experiment [14] and this is simply due to the break of the spatial symmetry of the tilted TEM$_{10}$ mode. The atom velocity is 0.42±0.01m/s. We have observed diversiform transmission spectrums. If y<0, the atom passes through the upper part of TEM$_{10}$ mode firstly and then passes through the lower part (see Fig. 4 (d)), and vise versa (see Fig. 4 (f)). Both of the atom trajectory and atom velocity are uniquely determined with this tilted higher order transverse mode in the high-finesse cavity. The precision is mainly limited by the horizontal movement during the transit process caused by the probe light and the uncertainty from the fluctuation of the Poisson photon counting statistics. Of cause, here we have one fundamental ambiguity: we don't know the specific antinode in which the atom passes through.

For a small fraction of atom transits, corresponding to those atoms passing away from the antinode and very close to the node of the cavity mode, the method of determining the off-axis *y* cannot be applied. For these atom transits, the transmission can not drop to the bottom and the trajectory determination fails in this case because we can not distinguish these atoms from those atoms passing through the antinode ($z\sim 0$) but far away from the cavity axis. For example, transits at $z$=200nm, $y$=0μm and $z$=0, $y$=36.5μm have almost the same transmission spectrums and we could not distinguish them experimentally. However, in the relative wide range of $z$=-150nm to +150nm, the transmission drops down to almost zero and what we have observed is most likely in this range.

We have also measured the average velocities of atoms by counting the different arrival times of the individual atoms from MOT to the cavity mode. The result is shown in Fig. 5. We find that there exists a minimum velocity at the arrival time of about 32ms, which corresponds to the time interval of an atom with zero velocity falls down freely in 5mm and it is coincident with the experiment. Those earlier and later arrived atoms with non-zero velocities, corresponding to downward and upward movement initially, have higher velocities. We thus observed a V-shape distribution of the velocity of the atoms and this velocity distribution is also coincident with the results in Fig.4. This gives the temperature of the atoms in the magneto-optic trap, 186μK±19μK.

In summary, we have investigated the trajectories and velocities of single neutral cesium atoms coupled to a tilted $TEM_{10}$ mode in a high-finesse optical cavity. The degenerate trajectory of single atom is eliminated completely due to the beak of the geometry of the spatial mode in the cavity. The unique track of the single atom is determined. The velocity of the atoms is also measured. The transmission spectrums are fitted very well in the weak-field approximation of the strong coupling between a tilted $TEM_{10}$ Hermite-Gaussian transverse mode and an atom. We yield 5.6μm of the spatial resolution in 10μs time interval along vertical direction (axis $x$) and 0.1μm for off-axis ($y$ direction). The result of this high resolution of off-axis benefits from the break of asymmetry of the tilted intra-cavity $TEM_{10}$ mode. The unique determination of the single atom trajectory is impossible for the usual $TEM_{00}$ mode. This method can be extended to sense the atomic center-of-mass motion and the kinetic energy of single atom in real time, with better spatial and time resolution when the higher transverse modes and smaller mode volume are used and this may develop the time-resolved microscopy based on the strong coupling atom-cavity microscope [14, 30].

The work is supported by the National Nature Science Foundation of China (Project No. 10794125).

**Figures Captions**

Fig. 1 (Color online) Theoretical Model. (The atoms fall down freely from MOT and pass through the cavity higher transverse modes. The optimum coupling constant $g_0$ is larger than cavity decay rate $\kappa$ and atom decay rate $\gamma$, so the interaction between atom

and cavity mode reaches the strong coupling regime)

Fig. 2 (Color online) Cavity transmission spectrums versus atom's positions in $x$ axis for $TEM_{10}$ mode. (a) The cavity transmission spectrum with different off-axis distances ($y=0, \pm20, \pm35, \pm40, \pm45\mu m$) between atoms' trajectories and the center of mode. The detuning between probe light and atom transition is 0MHz. (b) The cavity transmission spectrum with different detunings ($\Delta_{pa}= 2\pi\times(0, -10, -23.9)$MHz) for $y=0$.

Fig. 3 (Color online) Cavity transmission versus the scanning of the probe frequency. The inset shows $TEM_{00}$ mode and $TEM_{10/01}$ modes. Splitting about 82.5MHz between $TEM_{10}$ and $TEM_{01}$ mode is observed. Three images are the 2D intensity distributions of $TEM_{00}$, $TEM_{10}$ and $TEM_{01}$ modes directly detected by a CCD camera.

Fig. 4 (Color online) The cavity transmission spectrums of a single atom coupled to the tilted $TEM_{10}$ mode of a high-finesse optical cavity. Red dots are the experimental data and green solid curves are theoretical fitting according to the expression (3). (a) $y=-16.3\pm0.1\mu m$, $v=0.39\pm0.01$m/s. (b) $y=0\pm0.2\mu m$, $v=0.42\pm0.01$m/s. (c) $y=18\pm0.1\mu m$, $v=0.42\pm0.01$m/s. The right side figures (d), (e) and (f) show the unique atom trajectories corresponding to (a), (b) and (c), respectively. $\Delta_{pa}=0$MHz. [$g_0=2\pi\times23.9$MHz, $w_0=23.8\mu m$, $\gamma=2\pi\times2.6$MHz and $\kappa=2\pi\times2.6$MHz]

Fig. 5 (Color online) The velocities of atoms at the cavity at different arrival times.

**Fig. 1**

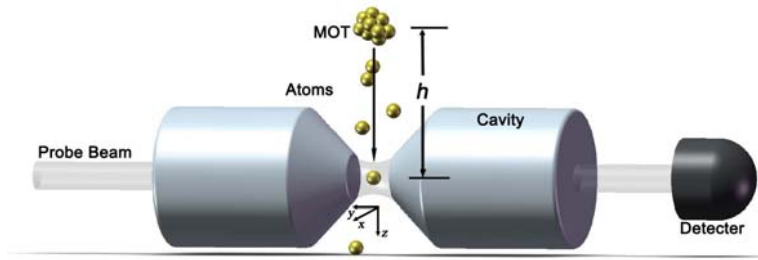

**Fig. 2**

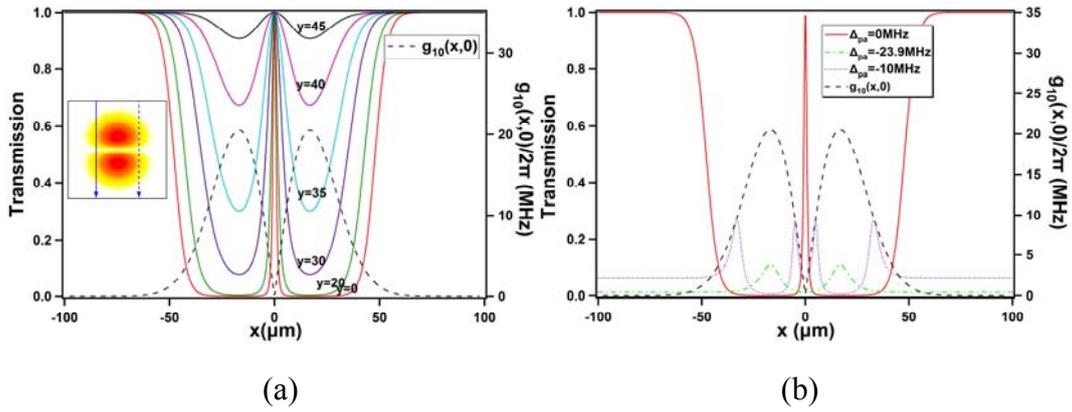

**Fig. 3**

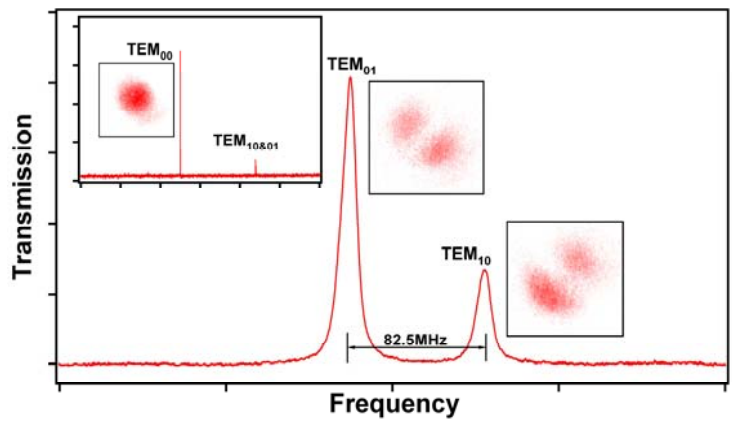

**Fig. 4**

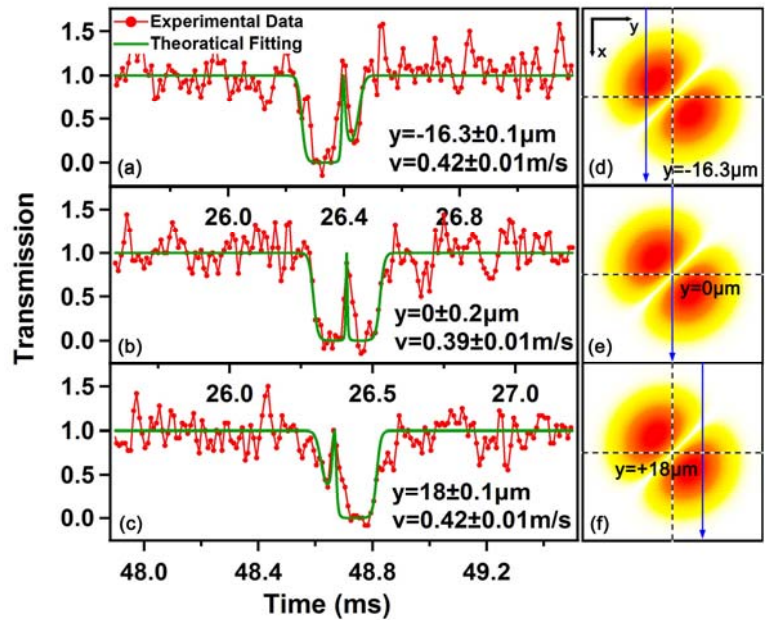

**Fig. 5**

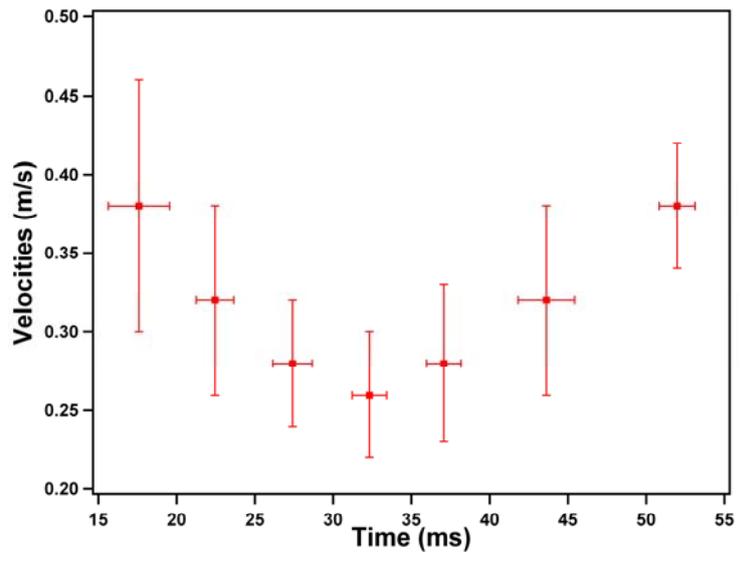